\documentclass[12pt,reqno]{amsart}
\pdfoutput=1 
\usepackage{amsmath,bbm}
\usepackage{latexsym}
\usepackage{amsfonts}
\usepackage{amssymb}
\usepackage{color}
\usepackage{graphicx}
\usepackage{url}
\usepackage{enumerate}
\usepackage{tikz}
\usetikzlibrary{shadings, intersections, calc, plotmarks}
\usepackage{marginnote}
\usepackage{stackrel}

\usepackage{geometry}
\xdefinecolor{tumblue}     {RGB}{0,101,189}
\xdefinecolor{tumgreen}    {RGB}{162,173,  0}
\xdefinecolor{tumred}      {RGB}{229, 52, 24}
\xdefinecolor{tumivory}    {RGB}{218,215,203}
\xdefinecolor{tumorange}   {RGB}{227,114, 34}
\xdefinecolor{tumlightblue}{RGB}{152,198,234}

\newtheorem{theorem}{Theorem}
\newtheorem*{theorem*}{Theorem}
\newtheorem{lemma}{Lemma}
\newtheorem{corollary}{Corollary}
\newtheorem*{corollary*}{Corollary}

\newcommand{\R}{\mathbbm{R}}
\newcommand{\C}{\mathbbm{C}}
\newcommand{\N}{\mathbbm{N}}

\newcommand{\F}{\mathbbm{F}}
\newcommand{\Q}{\mathbbm{Q}}

\newcommand{\1}{\mathbbm{1}}

\def\>{{\rangle}}
\def\<{{\langle}}
\def\Z{{\mathbb{Z}}}
\newcommand{\be}{\begin{equation}}
\newcommand{\ee}{\end{equation}}
\newcommand{\bea}{\begin{eqnarray}}
\newcommand{\eea}{\end{eqnarray}}

\newcommand{\tr}[1]{\mathrm{tr}\left[#1\right]} 

\newcommand{\norm}[1]{\left\lVert #1 \right\rVert}

\newcommand{\margintxt}[1]{}

\begin{document}

\title[Transcendental properties of entropy-constrained sets]{ Transcendental properties of entropy-constrained sets}

\author[Blakaj]{Vjosa Blakaj$^{1,2}$}
\email{vjosa.blakaj@tum.de}
\author[Wolf]{Michael M. Wolf$^{1,2}$}
\email{m.wolf@tum.de}
\address{$^1$ Department of Mathematics, Technical University of Munich}
\address{$^2$ Munich Center for Quantum
Science and Technology (MCQST),  M\"unchen, Germany}

\begin{abstract} For information-theoretic quantities with an asymptotic operational characterization, the question arises whether an alternative single-shot characterization exists, possibly including an optimization over an ancilla system. If the expressions are algebraic and the ancilla is finite, this leads to semialgebraic level sets. In this work, we provide a criterion for disproving that a set is semialgebraic based on an analytic continuation of the Gauss map. Applied to the von Neumann entropy, this shows that its level sets are nowhere semialgebraic in dimension $d\geq 3$, ruling out algebraic single-shot characterizations with finite ancilla (e.g., via catalytic transformations). We show similar results for related quantities, including the relative entropy, and discuss under which conditions entropy values are transcendental, algebraic, or rational.
\end{abstract}

\maketitle
\tableofcontents

\section{Introduction}\label{sec:intro}

\emph{Semialgebraic sets} are ubiquitous in quantum information theory. This is to some extent due to the fact that a large part of the theory is formulated in finite-dimensional vector spaces and that many often used constraints (like positivity of operators) are semialgebraic. 
Another reason for the prolificness of semialgebraic sets are diverse closure properties of such sets — especially the Tarski-Seidenberg theorem (1.4 and 2.2 in \cite{RAG98}), which shows that  arbitrary quantifiers ($\exists$, $\forall$) over sets that are themselves semialgebraic can be used without leaving the semialgebraic world. This world can be left, however, when using non-algebraic functions or limits, for instance those related to  unbounded dimensions or to an asymptotic number of copies in information theoretic contexts. To put it boldly, quantification over an integer variable is the nemesis of the semialgebraic world.\vspace*{5pt}

The present work is devoted to deciding whether or not certain sets are semialgebraic. We will first derive a general criterion that is especially suited  for  sets that are implicitly defined as preimages of functions and then apply it to the level sets of the entropy function and related quantities.\vspace*{5pt}

We want to begin, however, with a motivating example in which the question `semialgebraic or not?’ arises rather naturally. Consider the following case of catalytic state transformations, which was studied in \cite{CatalyticEntropy, RevCatalysis}: 

For an initial state $\rho’$ on $\C^{d}$, define $\mathcal{S}_n$ as the set of all states $\rho$ on $\C^{d}$ with the property that for any $\epsilon>0$ there is a state $\sigma$ on $\C^{n}$ and a unitary $U$ on $\C^d\otimes\C^n$ such that the reduced states of $\rho_{12}:=U(\rho’\otimes\sigma)U^*$ satisfy $\rho_2=\sigma$ and $\|\rho_1-\rho \|_1\leq\epsilon$. In other words,  $\mathcal{S}_n$ is the set of states that can be reached (approximately) from $\rho’$ with the help of an $n$-dimensional `catalyst’. As a second set, consider the set $\mathcal{S}$ of all states $\rho$ on $\C^{d}$ whose entropy is larger or equal to the one of $\rho’$. It follows from the subadditivity of the entropy \cite{CatalyticEntropy, RevCatalysis} that
\begin{equation}\label{eq:catalinclusion}
    \mathcal{S}_n\subseteq \mathcal{S},
\end{equation}
and the question arises whether equality holds in Eq.(\ref{eq:catalinclusion}) for some $n$ depending on $d$. There are various ways of addressing this question and coming to the conclusion that this is not the case despite the fact that $\mathcal{S}_\infty=\mathcal{S}$ \cite{RevCatalysis}. The arguably simplest one would be to exclude equality by arguing that $\mathcal{S}_n$ is semialgebraic (courtesy of Tarski-Seidenberg) while $\mathcal{S}$ is not. After all, the definition of $\mathcal{S}$ involves the logarithm, which is the paradigm of a non-algebraic function. While, as we see later, this way of reasoning is essentially correct, a more careful argument is required. Consider for instance $d=2$. In this case, $\mathcal{S}$ has an alternative semialgebraic characterization as the set of states for which $\tr{\rho^2}\leq c$ for a suitable constant $c$. Hence, for $d=2$ entropy-constrained sets \emph{are} semialgebraic and the transcendental nature of the logarithm only becomes relevant in the interrelation (and not in the intra-relation) of level sets. As we will show below, this changes when $d>2$ where $\mathcal{S}$ becomes indeed transcendental. This does not only rule out equality in Eq.(\ref{eq:catalinclusion}) but any kind of semialgebraic characterization of entropy constrained sets.

The use of transcendentality as a simple argument for ruling out specific representations has a long but difficult to trace history. Here we want to mention at least \cite{FCS}, where it was shown that the ground state of the antiferromagnetic spin 1/2 Heisenberg chain cannot be a finitely correlated state, since its energy is transcendental.

An interesting problem that is at least superficially related to the one we consider is the question whether \emph{almost-entropic regions}, which are specified by linear information inequalities for the Shannon entropy, are semialgebraic \cite{AlmostEntropicRegion}.

\section{Preliminaries}\label{sec:prelim}
We call a function $f:\R^n\rightarrow\R^k$ \emph{algebraic} over a subfield $\F\subseteq\R$ if for each of its $k$ component functions $f_i$ there is a polynomial $p_i\in \F[y,x_1,\ldots, x_n]$ such that $y=f_i(x)\Leftrightarrow p_i(y,x_1,\ldots,x_n)=0$. We will usually apply this concept only locally, i.e., to a neighborhood of a point on the graph of $f$.

By $H_d\subseteq\C^{d\times d}$ we denote the space of Hermitian $d\times d$ matrices and by $P_d\subseteq H_d$ the set of positive definite matrices with non-degenerate spectrum. The subset of trace-one matrices will be denoted by $D_d \subseteq P_d$. That is, $D_d$ is the set of density matrices that are non-degenerate and have full rank. We can identify $H_d$ with $\R^{d^2}$ by collecting all relevant real and imaginary parts of matrix entries into one vector. The resulting map is then a vector-space isomorphism and we write $\nu:\R^{d^2}\rightarrow H_d$ for its inverse. $P_d$ is an open subset of $H_d$ with respect to the usual topology that is induced by any norm on $H_d$. Accordingly, $\nu^{-1}(P_d)$  becomes an open subset of $\R^{d^2}$ and thus a smooth $d^2$-dimensional submanifold. A function $f:H_d\rightarrow\R^k$ will be called algebraic, if $f\circ\nu$ is algebraic. 

A subset of  $\R^n$ is called semialgebraic if it can be defined by a finite number of polynomial equations and inequalities. Unless otherwise stated, all the involved polynomials are over $\R$. A map between semialgebraic sets is called semialgebraic if its graph is a semialgebraic set.
\section{Properties of semialgebraic sets}\label{sec:semialg}

In this section we will have a closer look at general properties of semialgebraic sets---aiming at criteria that allow us to show that certain subsets of a Euclidean space are not semialgebraic.

Every semialgebraic set $S\subseteq\R^n$ is the disjoint union 
\begin{equation}\label{eq:stratification}
\dot\bigcup_{i=0}^p M_i=S
\end{equation} of finitely many \emph{Nash submanifolds} of $\R^n$, i.e.,  submanifolds  that are at once smooth and semialgebraic (see $2.3$ in \cite{IntroSemiAlgGeo}).  We can choose this decomposition (a.k.a. \emph{stratification}) such that each $M_i$ is connected but we can also choose it such that $i$ is the manifold-dimension of $M_i$.  
A \emph{Nash map} between Nash submanifolds of Euclidean spaces is a map that is both smooth and semialgebraic. 
For more details on semialgebraic sets, Nash manifolds, and Nash maps we refer to \cite{RAG98,Finitenash}. 

Denoting by $N_x M_i$  the normal space of $M_i$ at $x\in M_i$, we  define the \emph{normal bundle} of $S$ as
$$NS:=\bigcup_{i=0}^p\big\{(x,y)|x\in M_i,\; y\in N_x M_i\big\} \subseteq\R^n \times \R^n . $$

In the following, our main focus will be on the so-called \emph{Gauss map}, which maps a point on a manifold to its normal space. One reason behind this is that if the manifold is implicitly defined as the preimage of some function (such as the level sets of the entropy), the Gauss map is often easier to handle than the manifold itself. Moreover, it has the following simple but crucial property:

\begin{lemma}\label{lem:semialgNS}
If $S\subseteq\R^n$ is semialgebraic, then for any corresponding Nash submanifold $M_i\subseteq S$ the map from $x\in M_i$ to the orthogonal projector $P(x)$ onto $N_x M_i$ is a Nash map.  In particular, the normal bundle $NS$ is semialgebraic.
\end{lemma}
\begin{proof}
Consider the squared distance function $\eta:\R^n\ni x\mapsto\frac12\inf\{\|x-y\|^2 | y\in M_i\} $.  
Since $M_i$ is a smooth submanifold of $\R^n$,  according to Thm.3.1 in \cite{AmbrosioSoner96}, the Hessian $\nabla^2\eta(x)$ represents the orthogonal projection $P(x)$ onto the normal space $N_x M_i$ and it depends smoothly on $x\in M_i$.  Moreover, as $M_i$ is in addition semialgebraic,  Prop.2.2.8 in \cite{RAG98} implies that $\eta$ is semialgebraic.  Since derivatives  of semialgebraic functions remain semialgebraic (Sec.4 in \cite{JohnNixon} or Prop.2.9.1. in \cite{RAG98}),  $x\mapsto \nabla^2\eta(x)=P(x)$ is semialgebraic and smooth, hence a Nash map.  The normal bundle $NS$ is a  union of $p$ graphs of such maps and therefore a semialgebraic set.
\end{proof}

\begin{theorem}\label{thm:semanalytic} 
Let $M$ be a Nash submanifold of $\R^n$, and $P(x)\in \R^{n\times n}$ the orthogonal projector onto the normal space of $M$ at $x\in M$. For any pair of Nash maps $g:I\subseteq\R\rightarrow M$ and $h:\R^{n\times n}\rightarrow\R$ define $f:=h\circ P\circ g:I\rightarrow\R$. Then \begin{enumerate}
\item $f$ is analytic and algebraic over $\R$,
\item the global analytic function obtained from $f$ by analytic continuation has a compact Riemann surface and, in particular,  a finite number of branches. 
\end{enumerate}
\end{theorem}
\begin{proof}
Lemma \ref{lem:semialgNS} implies that $f$ is a composition of Nash maps and thus itself a Nash map \footnote{Since the composition of smooth maps is smooth, and the composition of semialgebraic maps is semialgebraic (Prop.2.2.6 \cite{RAG98}), it follows that the composition of Nash maps is a Nash map.}. Prop.8.1.8 in \cite{RAG98} states that a function from a semialgebraic set into $\R$, like $f$,  is  a Nash map if and only if it is analytic and algebraic over $\R$. This proves $(1)$.

In order to arrive at (2) we use that any real analytic function $f$ on an interval $I$ can be extended uniquely to a complex analytic function in an open neighborhood of $I\times\{i0\}\subseteq\C$ and further to a global analytic function by analytic continuation.   Regarding the polynomial $p\in\R[y,x]$ that governs the algebraic relation $p(f(x),x)=0$ as an element in  $\C[y,x]$,  the complex analytic extension of $f$ is still algebraic: the same polynomial relation interpreted over $\C$ locally defines a unique function (\cite{Ahlfors}, Chap. 8) that has to coincide with the (extension of) $f$ by the identity principle.  Algebraic analytic functions are known to have  at most a finite number of branches (Thm.4, p.306 in \cite{Ahlfors}) and a compact Riemann surface (I.\S 2 in \cite{AlgebraicCurves}).
\end{proof}
In order to show transcendentality of entropy-constrained sets, the idea is to use the last part of (2) in Thm.\ref{thm:semanalytic} and to seek a contradiction to the fact that the logarithm-function has an infinite number of branches.
\section{Transcendental entropy values}\label{sec:result}
Before we discuss transcendental sets and functions, we will make a brief excursion to the related topic of transcendental numbers and analyze under which conditions the entropy of a single state is transcendental, algebraic or rational. $\overline{\Q}$ will denote the field of algebraic numbers. 
The main tool from transcendental number theory on which our results are based on is:
\begin{lemma}[Baker's theorem \cite{Baker}]\label{thm:Baker} Let $\lambda_1,\ldots,\lambda_n\in\overline{\Q}\setminus\{0,1\}$ be such that $\ln \lambda_1,\ldots,\ln \lambda_n$ are linearly independent over $\Q$. Then $1, \ln \lambda_1,\ldots,\ln \lambda_n$  are linearly independent over $\overline{\Q}$.
\end{lemma}
Equipped with this instrument we obtain the following dichotomies:            
\begin{theorem}\label{thm:Bakerconsequence}
Let $\rho\in\C^{d\times d}$ be a density matrix and $S(\rho):=-\tr{\rho\log_b\rho}$ where $b>1$ is the base of the logarithm.\begin{enumerate}
    \item If $b=e$ (i.e. $\log_b=\ln$) and the eigenvalues of $\rho$ are algebraic, then $S(\rho)$ is either zero or transcendental.
    \item If $b$ is algebraic and the eigenvalues of $\rho$ are rational, then $S(\rho)$ is either rational or transcendental.
\end{enumerate} 
\end{theorem}

\emph{Remark: } Note that the eigenvalues are in particular algebraic if the matrix elements of $\rho$ are, since then $\det(\lambda\1-\rho)=0$ becomes a polynomial in $\overline{\Q}[\lambda]$. 
\begin{proof} Let $\lambda_1,\ldots,\lambda_d$ be the eigenvalues of $\rho$.

(1): A consequence of Lem.\ref{thm:Baker} is that, under the assumptions of the Lemma, any $\overline{\Q}$-linear combination of $\ln \lambda_1,\ldots,\ln \lambda_n$ is either zero or transcendental. This can be seen by induction over $n$. For $n=1$ it follows immediately from Lem.\ref{thm:Baker}. The induction step from $n$ to $n+1$ can be shown by contradiction: suppose some $\overline{\Q}$-linear combination  \begin{equation}\label{eq:Bakerentropytranscendental}
    \beta:=\sum_{i=1}^{n+1}\alpha_i\ln \lambda_i ,\quad \alpha_i\in\overline{\Q}
\end{equation} is a non-zero algebraic number $\beta$. Then by Lem.\ref{thm:Baker} one of the $\ln \lambda_i$ is a $\Q$-linear combination of the other $n$ and can thus be replaced by them in Eq.(\ref{eq:Bakerentropytranscendental}), which would contradict the induction hypothesis. 

(2): If we define $a:=\prod_i \lambda_i^{-\lambda_i}$ where the product runs over all non-zero eigenvalues, then by assumption $a\in\overline{\Q}$ and  $S(\rho)=(\ln a)/(\ln b)$. Hence, if $S(\rho)\in\overline{\Q}$, then $\ln a$ and $\ln b$ would be $\overline{\Q}$-linearly dependent. By Lem.\ref{thm:Baker} this would imply that they are $\Q$-linearly dependent, which in turn implies that their fraction $S(\rho)$ must be rational.
\end{proof}
A simple consequence of Thm.\ref{thm:Bakerconsequence} is that when using the natural logarithm, level sets of the entropy cannot be semialgebraic over $\overline{\mathbbm{Q}}$ unless the entropy is zero. Next, we show that something similar is true over $\R$.

\section{Entropy-surfaces are nowhere semialgebraic}

In this section, we present the application of Thm.\ref{thm:semanalytic} to von Neumann entropy level sets. The sets of states that have extremal entropy ($0$ or $\ln d$) are evidently (semi)algebraic in any dimension $d$. Moreover, as we have discussed in the introduction, all level sets of the entropy are semialgebraic for $d=2$. The following shows that this is no longer the case for $d>2$.
\begin{theorem}\label{thm:main2}
For any $d\geq 3$ and  $c\in(0,\ln d)$ the set of $d\times d$ density operators whose von Neumann entropy is equal to $c$ is nowhere semialgebraic. That is, if  ${\mathcal S}:=\{\rho\in H_d|\rho\geq0,\tr{\rho}=1,-\tr{\rho\ln\rho}=c\}$, then for any open set $V\subset H_d$ the set $\mathcal{S}\cap V$ is not semialgebraic unless it is empty. 
\end{theorem}
\emph{Remarks:} 1. Theorem and proof use the natural logarithm. However, the same statement holds true as well for any other base, since changing the base of the logarithm is equivalent to changing the value of $c$. 

2. Here and in the following section, we only consider the case of equality ``$=c$''. Since boundaries of semialgebraic sets are semialgebraic, the same result holds true for the inequalities ``$<c$'', ``$\leq c$'', ``$>c$'' and ``$\geq c$''.

\begin{proof}
Since we consider neighborhoods of states of constant entropy, we can restrict ourselves to sufficiently small subsets and thereby assume w.l.o.g. that $(\mathcal{S}\cap V)\subset D_d$. In this way, we can regard $\mathcal{S}\cap V$ as a smooth $(d^2-2)$-dimensional submanifold of $D_d$,  since any $c\in(0,\ln d)$ is a regular value of the entropy function. 

Next, we employ a map $\Phi:D_d\rightarrow\R^{d-1}\times\R^{d^2-d}$ that maps any $\rho\in D_d$ onto a vector whose first $d-1$ components are distinct eigenvalues of $\rho$. As proven in Lem.\ref{lemma:Phi} in  Appendix \ref{app:diff}, we can choose $\Phi$ such that it becomes an algebraic diffeomorphism onto its range when restricted to a sufficiently small neighborhood. Consequently, if $(\mathcal{S}\cap V)$ is semialgebraic, then its image $M:=\Phi(\mathcal{S}\cap V)$ would be a Nash submanifold of $\R^n$, $n:={d^2-1}$ to which Thm.\ref{thm:semanalytic} applied. We now assume that this is the case and seek for a contradiction. 

Let $\rho$ be any state in $\mathcal{S}\cap V$.
The manifold $M$ is the preimage of $c$ under the map $F:\R^n\rightarrow\R$,
$$ F(x):= -\Big(1-\sum_{i=1}^{d-1} x_i\Big)\ln \Big(1-\sum_{i=1}^{d-1} x_i\Big) - \sum_{i=1}^{d-1} x_i\ln x_i, $$ in a neighborhood of $\Phi(\rho)$.

\begin{figure}[t]
    \centering
    \includegraphics[width=1\textwidth]{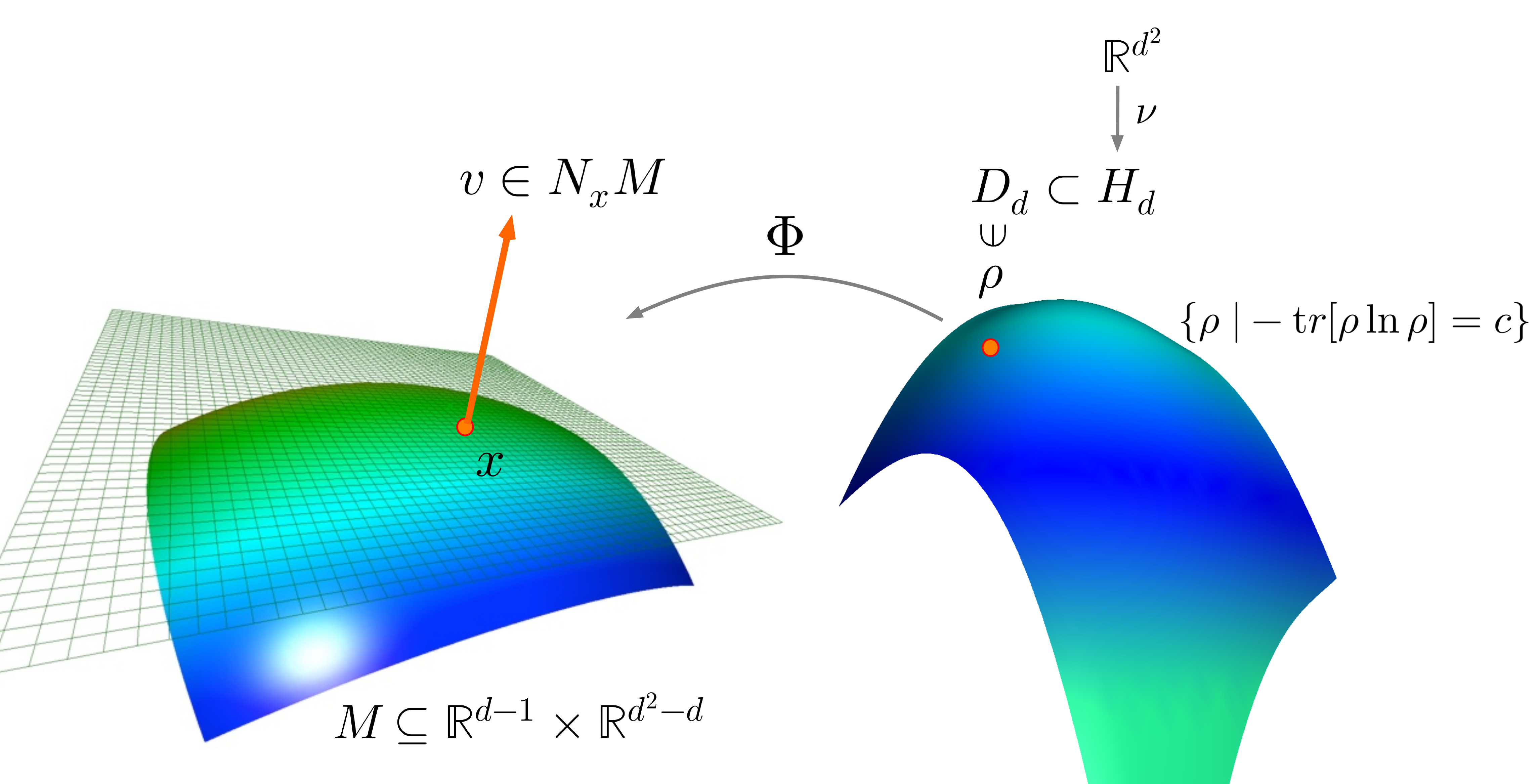} 
    \caption{Visualization for Lem.\ref{lemma:Psi}, Lem.\ref{lemma:Phi} and the proof of Thm.\ref{thm:main2}. $M$ is the considered manifold that is obtained via the diagonalizing algebraic diffeomorphism $\Phi$ acting locally on the set of states of constant entropy.} \label{fig:dimklemma}
\end{figure}

Standard results in differential geometry tell us that the normal space of $M$ at $x\in M$ is the one-dimensional space spanned by the gradient $\nabla F(x)$. The components of the gradient are $\nabla F(x)_i=\ln\big(1-\sum_{i=j}^{d-1} x_j\big)-\ln x_i$ for $i<d$ and zero otherwise. In order to apply Thm.\ref{thm:semanalytic} we define a Nash map $h:\R^{n\times n}\rightarrow\R$, $h(X):=\sqrt{X_{11}/X_{22}}$. Composing this with the projector $P(x)$ onto the normal space we obtain (under the assumption that $d\geq 3$):
\begin{equation}\label{eq:hPx}
    h\circ P(x) = \left|\frac{\nabla F(x)_1}{\nabla F(x)_2}\right|.
\end{equation}
We can assume that the components of $x$, i.e. the eigenvalues of $\Phi^{-1}(x)$, are in ascending order for all $x$ in the considered neighborhood. In this way, the absolute value in Eq.(\ref{eq:hPx}) can be neglected, as the quotient is positive.

Now consider a path on $M$ parameterized by $g:\R\supset I\rightarrow M$ that goes through $\xi:=\Phi(\rho)$ so that $g(\lambda_1):=(\lambda_1,\lambda_2,\xi_3,\ldots,\xi_n)$ where $\lambda_2=\lambda_2(\lambda_1)$ is implicitly defined by demanding $g(I)\subset M$. Here, the implicit function theorem guarantees the existence of $\lambda_2(\lambda_1)$ as solution to $F(\lambda_1,\lambda_2,\xi_3,\ldots,\xi_n)=c$.
If $M$ is a Nash manifold, then $\lambda_1\mapsto\lambda_2(\lambda_1)$ is algebraic and $g$ is a Nash map on a neighborhood of $\xi_1$ at which $g(\xi_1)=\Phi(\rho)$. 

According to Thm.\ref{thm:semanalytic} the analytic continuation of the function
\begin{equation}\label{eq:fhpg}
    f(\lambda_1):=h\circ P\circ g(\lambda_1)=\frac{\ln\left(\frac{\lambda_1}{w-\lambda_1-\lambda_2(\lambda_1)}\right)}{\ln\left(\frac{\lambda_2(\lambda_1)}{w-\lambda_1-\lambda_2(\lambda_1)}\right)},\quad\text{where}\quad w:=1-\sum_{j=3}^{d-1}\xi_j
\end{equation}
must lead to a global analytic function with only finitely many branches---if the hypothesis of $M$ being semialgebraic is valid. In the remaining part of the proof, we will show that this is not that case and that $f$ gives rise to an infinite number of branches.

 Let $z,y: I \subseteq \R \longrightarrow \R$ denote the functions
$z(\lambda_1):= \frac{\lambda_1}{w-\lambda_1-\lambda_2(\lambda_1)}$ and $y(\lambda_1):=\frac{\lambda_2(\lambda_1)}{w-\lambda_1-\lambda_2(\lambda_1)}$, respectively. Under the assumption that  $\lambda_1\mapsto\lambda_2(\lambda_1)$ is an algebraic function, and using the closure properties of the set of algebraic functions (Sec.4 in \cite{JohnNixon}, Thm.6.4 in \cite{AVNI2020107241}), $z$ and $y$ are  algebraic functions as well. As such, they can be regarded as global analytic functions defined on the entire complex plane up to a finite number of points (Thm.3.1, \cite{NonAlgebraicFunction}) and with at most finitely many branches. 

Consider the analytic continuation of $f$ along a closed path $\gamma$ in the complex plane that starts at $\xi_1$, bypasses all of the finitely many singularities and is such that the image $z(\gamma)$ goes around the origin once and returns to the initial function value. Since algebraic functions have algebraic inverses, such a path always exists. As $y$ has a finite number of branches, there is a $k\in\N$ such that also $y$ returns to the same function value if we run through $\gamma$ $k$-times in a row. After continuously tracing the path $\gamma$ $nk$ times for any $n\in\Z$, the branches of the complex logarithm give rise to a change of the value of the function $f$ from the initial $f(\xi_1)$ to 
\begin{equation}\label{eq:fvalues}
 \frac{2\pi i k n+\ln z(\xi_1)}{2\pi i m(n) +\ln y(\xi_1)},
\end{equation}
where $m:\Z\rightarrow\Z$ is some function that takes into account how many times the image of the path under $y$ has enclosed the origin. As shown in Lem.\ref{lem:infinity} in Appendix \ref{app:tech}, if we run over all $n\in\Z$, then Eq.(\ref{eq:fvalues}) represents an infinite number of function values as long as $f(\xi_1)$ is irrational. In case $f(\xi_1)$ happens to be rational, we apply the same argument albeit with the starting point $\xi_1$ slightly shifted to a point $\Tilde{\xi}_1\in I$ for which $f(\Tilde{\xi}_1)\not\in\Q$. That such a $\Tilde{\xi}_1$ exists in any neighborhood of $\xi_1$ is implied by the fact that $f$ is continuous and non-constant, since $f'(\xi_1)\neq 0$, which is proven in Lem.\ref{lem:fnonconstant} in Appendix \ref{app:tech}. 
\end{proof}

From the proof of the above theorem, we can see that the same outcome holds true for the classical case where the Shannon entropy is used instead of the von Neumann entropy.

\section{Additional constraints and relative entropy}

In this section, we want to extend the result by first allowing for an additional constraint and then showing a similar result for the relative entropy. Additional constraints could concern the distance of the state to a specific target, its energy or the expectation value of any observable. Mathematically, we will describe those using a differentiable function $h:P_d\rightarrow\R$ for which we denote by $\nabla h(\rho)\in H_d$ the gradient, i.e., the operator that is related to the Fréchet derivative $h'(\rho)$ via $h'(\rho):H_d\ni X\mapsto\tr{X\nabla h(\rho)}$.

\begin{theorem}\label{thm:plusconstraint}
For $d\geq 3$, $c_1\in(0,\ln d)$, $c_2\in\R$ and $h\in C^1(P_d,\R)$ let $\rho$ be an element of $$ {\mathcal S}:=\{\rho\in P_d\;|\;\tr{\rho}=1,-\tr{\rho\ln\rho}=c_1, h(\rho)=c_2\},$$ with $[\rho,\nabla h(\rho)]\neq 0$. Then $\mathcal{S}$ is not semialgebraic in any neighborhood of $\rho$.
\end{theorem}

\emph{Remark:} The condition $[\rho,\nabla h(\rho)]\neq 0$ is stronger than necessary but conveniently serves the purpose of the proof. It can be interpreted as a first-order-way of imposing that $h(\rho)$ does not solely depend on the spectrum of $\rho$.\\

\begin{proof} Without loss of generality, we can consider a sufficiently small neighborhood in which $[\rho,\nabla h(\rho)]\neq 0$ holds for each of its elements.
We begin with convincing ourselves that within such a neighborhood $\mathcal{S}$ is a regular $C^1$-submanifold of $P_d$. To this end, we  regard $\mathcal{S}$ as the preimage of $c:=(1,c_1,c_2)$ under  $f:P_d\rightarrow\R^3$, $f(\rho):=(\tr{\rho},S(\rho),h(\rho))$. Then $c$ is a regular value, and thus $\mathcal{S}$ a regular $C^1$-submanifold, if the three involved gradients $\1,\nabla h(\rho)$ and $\nabla S(\rho)=-\1-\ln\rho$ (Lem.VI.9 in \cite{HansonDatta}) are linearly independent. This is, however, guaranteed by $[\rho,\nabla h(\rho)]\neq 0$.

Next, we adopt the viewpoint of $\mathcal{S}$ as an intersection of two manifolds, namely the level set of the von Neumann entropy studied in Thm.\ref{thm:main2} and the manifold $h^{-1}(\{c_2\})$. More precisely, we will consider a submersion of those manifolds into a space that is relevant for the argument in the proof of Thm.\ref{thm:main2}. The considered submersion is given by $\psi:P_d\rightarrow\R^d$, $\psi_i(\rho):=\Psi_i(\rho)$, where $\Psi$ is the diagonalizing algebraic diffeomorphism from Lem.\ref{lemma:Psi}.  We claim for the tangent space of one of the submersed manifolds that
\begin{equation}\label{eq:tangentspaceisRd}
    T_{\psi(\rho)}\Big[\psi\circ h^{-1}\big(\{c_2\}\big)\Big]=\psi'(\rho)\;T_\rho\Big[h^{-1}\big(\{c_2\}\big)\Big] =\R^d.
\end{equation}
Before we prove Eq.(\ref{eq:tangentspaceisRd}), let us see why Eq.(\ref{eq:tangentspaceisRd}) completes the proof of the theorem: the tangent space $T_{\psi(\rho)}\big[\psi(\mathcal{S})\big]$ of the intersected manifold is equal to the intersection of the tangent spaces of the individual manifolds.\footnote{Strictly speaking, this requires that the manifolds intersect transversally, but as one of the tangent spaces is the entire space, the intersection is trivially transversal.} However, by Eq.(\ref{eq:tangentspaceisRd}), the tangent space associated to the additional constraint is the entire space $\R^d$ so that this `intersection' becomes void and $T_{\psi(\rho)}\big[\psi(\mathcal{S})\big]$ is, in fact, the tangent space of the manifold already studied in Thm.\ref{thm:main2}. Consequently, the same argument applies.

Now let us show Eq.(\ref{eq:tangentspaceisRd}). The first equality just uses the Fréchet derivative as pushforward between tangent spaces. For the second equality we exploit that $\psi_i'(\rho):X\mapsto \langle\varphi_i|X|\varphi_i\rangle$, where $\{|\varphi_i\rangle\}$ is the eigenbasis  of $\rho$ (see proof of Lem.\ref{lemma:Psi}). Moreover, $T_\rho\big[h^{-1}\big(\{c_2\}\big)\big]$ is the orthogonal complement of $\nabla h(\rho)$. Seeking for a contradiction, suppose the last equation in Eq.(\ref{eq:tangentspaceisRd}) does not hold. Then, there would exist a non-zero $b\in\R^d$ such that $\sum_i b_i  \langle\varphi_i|X|\varphi_i\rangle =0$ holds for any $X\perp\nabla h(\rho)$. In other words, $B:=\sum_i b_i|\varphi_i\rangle\langle\varphi_i|$ would have to be proportional to $\nabla h(\rho)$. This is impossible since $[\rho,B]=0$ whereas $[\rho,\nabla h(\rho)]\neq 0$. 
\end{proof}

As an application of Thm.\ref{thm:plusconstraint} let us show transcendentality of the level sets of the relative entropy. The \emph{relative entropy} of two density operators $\rho,\sigma$ on $\C^d$ is defined as $S(\rho|\!|\sigma):=\tr{\rho\ln\rho}-\tr{\rho\ln\sigma}$ whenever ${\rm supp}(\rho)\subseteq{\rm supp}(\sigma)$ and $S(\rho|\!|\sigma):=\infty$ otherwise.
\begin{corollary}
For any $c>0$, $d\geq 3$, any positive definite density matrix $\sigma\in H_d$ and any open subset $U\subseteq H_d$ the set \begin{equation}\label{eq:relentlevelset}
   \mathcal{R}:= \big\{\rho\in D_d\;|\; S(\rho|\!|\sigma)=c\big\}\cap U
\end{equation} is not semialgebraic in $H_d$ unless it is empty. 
\end{corollary}
\begin{proof}
We use that $S(\rho|\!|\sigma)=-S(\rho)-\tr{\rho\ln\sigma}$ and that $\rho\in D_d$ excludes the extremal  values $0$ and $\ln d$ of the von Neumann entropy. If $\sigma=\1/d$, then the set $\mathcal{R}$ reduces to a level set of the von Neumann entropy so that Thm.\ref{thm:main2} applies. If $\sigma\neq\1/d$, we claim that there is a $\tilde{\rho}\in\mathcal{R}$ that does not commute with $\sigma$ and thus $[\tilde{\rho},\ln\sigma]\neq 0$. 

In order to show this, note that $\mathcal{R}$ is a manifold of dimension $d^2-2$. The manifold of all density operators commuting with $\sigma\neq\1/d$, which we denote by $C_\sigma$, however, can be shown to have dimension at most $(d-1)^2$: since the commutant is a proper von Neumann subalgebra and since the largest such subalgebra is isomorphic to the matrix algebra ${\mathcal{M}}_{(d-1)}\oplus {\mathcal{M}}_1$ (cf. Thm.5.6 in \cite{Farenick}), the dimension of the submanifold of density matrices that are contained in this subalgebra is at most $(d-1)^2$. This implies that for all $d\geq 3$ we have ${\rm dim}(C_\sigma)<{\rm dim}(\mathcal{R})$ so that $\mathcal{R}$ must contain elements outside the commutant of $\sigma$. Let $\tilde{\rho} $ be one such element.

We restrict the focus to a neighborhood of $\tilde{\rho}$ in which no element commutes with $\sigma$.  If $\mathcal{R}$ were semialgebraic in that neighborhood, then $\mathcal{R}$ intersected with the affine space of all $\rho$ for which $\tr{(\rho-\tilde{\rho})\ln\sigma}=0$ would be semialgebraic as well. However, this intersection is covered by Thm.\ref{thm:plusconstraint} with $h(\rho):=\tr{\rho\ln\sigma}$ so that $\nabla h(\tilde{\rho})=\ln\sigma$ and therefore $[\tilde{\rho},\nabla h(\tilde{\rho})]\neq 0$.
\end{proof}


\emph{Acknowledgments:} MMW acknowledges funding by the Deutsche Forschungsgemeinschaft (DFG, German Research Foundation) under Germany's Excellence Strategy –  EXC-2111 – 390814868. VB acknowledges support by the International Max Planck Research School for Quantum Science and Technology at
the Max-Planck-Institute of Quantum Optics.

\appendix
\section{Diagonalizing algebraic diffeomorphisms}\label{app:diff}

In this appendix, we show that the diagonalization employed in the proof of Thm.\ref{thm:main2} can be done by means of an algebraic diffeomorphism.

\begin{lemma}\label{lemma:Psi} For any $\rho\in P_d$ there is an open neighborhood $U\subseteq P_d$ and a map $\Psi:U\rightarrow \R^{d^2}$ such that (i) $\Psi$ is algebraic over $\Q$, (ii) $\Psi$ is a diffeomorphism onto its range, and (iii) $\{\Psi_i(X)\}_{i=1}^d={\rm spec}(X)$ for any $X\in U$. 
\end{lemma}

\begin{proof}
As demanded by (iii),  we define $\Psi_i(X)$ to be the $i$'th eigenvalue of $X$ for $i=1,\ldots, d$.  This leads to algebraic functions over $\Q$ due to the polynomial relation ${\rm det}\big(\Psi_i(X)\1-X\big)=0$. 

From eigenvalue perturbation theory of Hermitian matrices (an adapted version of Thm.2.3 in \cite{MatrixTheory} where the left and right eigenvectors coincide) we know that  $\Psi_i(\rho+t A)=\Psi_i(\rho)+\tr{A |\varphi_i\rangle\langle\varphi_i|}t +{\mathcal{O}}(t^2)$ holds in the limit  $t\rightarrow 0$, where $|\varphi_i\rangle$ is the normalized eigenvector of $\rho$ corresponding to $\Psi_i(\rho)$. This implies that the derivative $\Psi_i'(\rho):H_d \rightarrow\R$ acts as $A\mapsto\tr{A|\varphi_i\rangle\langle\varphi_i|}$ and since the $\varphi_i$'s form an orthonormal basis, the $d$ derivatives $\Psi_1'(\rho),\ldots, \Psi_d'(\rho)$ are linearly independent.

To construct the remaining component functions $\Psi_i$ with $i>d$, consider the map $\Tilde{\Psi}:\R^{d^2}\rightarrow\R^d \times \R^{d^2}$, $$\Tilde{\Psi}(x):=\Big(\Psi_1\big(\nu(x)\big),\ldots,\Psi_d\big(\nu(x)\big),x \Big).$$
The Jacobi matrix $\Tilde{J}$ that represents the derivative of $\Tilde{\Psi}$ at $\nu^{-1}(\rho)$ has the form $\Tilde{J}={{C}\choose{\1}}$, where $C$ is a $d\times d^2$ matrix that represents the derivatives of the $d$ eigenvalues. As we have seen above, those are linearly independent, so that $C$ has rank $d$.  Hence, we can find $d$ rows within the $\1$-block of $\Tilde{J}$ that can be erased so that the resulting square matrix is non-singular. Denote this square matrix by $J$ and the map that selects the rows by $s:\R^d \times \R^{d^2}\rightarrow \R^{d^2}$ so that $J=s\Tilde{J}$. The derivative of the map $\Psi:=s\circ\Tilde{\Psi}\circ\nu^{-1}$ at $\rho$ then has full rank so that by the inverse function theorem there exists an open neighborhood  $U\ni\rho$ such that  $\Psi:U\rightarrow\Psi(U)$ is a diffeomorphism. 
\end{proof}
Based on this Lemma we can now show an analogous result that incorporates the constraint $\tr{\rho}=1$.
\begin{lemma}\label{lemma:Phi}
For any $\rho \in D_d$ there is an open neighborhood $V \subseteq D_d$ and a map $\Phi : V \longrightarrow \R^{d-1} \times \R^{d^2-d}$, such that $\Phi$ is (i) algebraic over $\Q$, (ii) a diffeomorphism onto its range and (iii) $\{\Phi(X)\}_{i=1}^{d-1}$ are distinct eigenvalues of $X$ for any  $X\in V$.
\end{lemma}
\begin{proof}
To construct such a map we compose the algebraic diffeomorphism $\Psi$ from Lem.\ref{lemma:Psi} with the following maps: $\iota : D_d \longrightarrow P_d$ denotes the inclusion map of the set $D_d$ into the larger set $P_d$ and $\pi : \R^d \times \R^{d^2-d} \longrightarrow \R^{d-1} \times \R^{d^2-d}$ the projection map that discards the $d$-th component of the input.
By $\Phi$ we denote the composition  $\Phi := \pi \circ \Psi \circ \iota : V \longrightarrow \R^{d-1} \times \R^{d^2-d}$. Being the composition of three smooth and algebraic maps, $\Phi$ is smooth and algebraic. It is also bijective onto its range with inverse $\Phi^{-1}=  \Psi^{-1} \circ \widehat{\pi}$, where 
$$\widehat{\pi}: (\lambda_1,..., \lambda_{d-1}, x) \mapsto \Big(\lambda_1,..., \lambda_{d-1}, 1-\sum_{j=1}^{d-1} \lambda_j, x\Big)\quad \text{with }x\in\R^{d^2-d}.$$ As a composition of smooth maps $\Phi^{-1}$ is smooth. Since a bijective smooth map with smooth inverse is a diffeomorphism, this concludes the proof.
\end{proof}

\section{Technical Lemmas}\label{app:tech}

\begin{lemma}\label{lem:infinity}
Let $a, b \in \R\setminus\{0\}$ be such that $\frac{a}{b}\not\in\Q$, $k \in \N$ and $m : \mathbb{Z} \longrightarrow \mathbb{Z}$. Then
$$\Big | \Big \{\frac{a + 2\pi ikn}{b + 2\pi i m(n)}\Big \}_{n \in \mathbb{Z}} \Big | = \infty.$$ 
\end{lemma}
\begin{proof}
Choose $n, \hat{n} \in \mathbb{Z}$ such that $n \neq \hat{n}$. Let $m := m(n)$ and $\hat{m} := m(\hat{n})$, and suppose that these give rise to the same value, i.e. 
\begin{equation}\label{eq:idvalues}
    \frac{a+2\pi i k n}{b + 2 \pi i m} = \frac{a + 2\pi i k \hat{n}}{b + 2\pi i \hat{m}}.
\end{equation}
By bringing together the imaginary parts we obtain from Eq.(\ref{eq:idvalues}) 
$$a (\hat{m}-m) + b k  (n - \hat{n}) = 0.$$
Since $n-\hat{n} \neq 0$ we see that this is only possible if $\frac{a}{b} \in \Q$. In other words, if $\frac{a}{b}\not\in\Q$, then different $n\in\Z$ lead to different values.\end{proof}
For the following Lemma and its proof we use the notation of the proof of Thm.\ref{thm:main2}.
\begin{lemma}\label{lem:fnonconstant}
The function $f:\R\supset I\rightarrow\R$ defined in Eq.(\ref{eq:fhpg}) satisfies $f'(\xi_1)\neq 0$.
\end{lemma}
\begin{proof}
With $C(\lambda_1,\lambda_2):=F(\lambda_1,\lambda_2,\xi_3,\ldots,\xi_n)$ the function $\lambda_2(\lambda_1)$ is implicitly defined as solution to $C\big(\lambda_1,\lambda_2(\lambda_1)\big)=c$. The implicit function theorem provides us with the derivative $\lambda_2'(\lambda_1)=-f(\lambda_1)$. Hence, $f'(\xi_1)=-\lambda_2''(\xi_1)$.
We once again invoke implicit differentiation in the form
\begin{equation}\label{eq:CHC}
    0\stackbin{!}{=}\frac{d^2}{(d\lambda_1)^2}C\big(\lambda_1,\lambda_2(\lambda_1)\big) = v^T H v+\partial_2 C\big(\lambda_1,\lambda_2(\lambda_1)\big)\lambda_2''(\lambda_1),
\end{equation}
with $v:=\big(1 \,\,\,  \lambda_2'(\lambda_1)\big)^T$ and the Hessian $H_{ij}:=\partial_i\partial_j C\big(\lambda_1,\lambda_2(\lambda_1)\big)$. The latter can be computed to $H_{ij}=-\delta_{ij}\lambda_i^{-1}-(w-\lambda_1-\lambda_2)^{-1}$ and is thus negative definite, which reflects the strict concavity of the entropy function. Therefore $v^T H v<0$ so that Eq.(\ref{eq:CHC}) implies $\lambda_2''(\lambda_1)\neq 0$ for any $\lambda_1$ in a neighborhood of $\xi_1$.

\end{proof}

\bibliographystyle{ieeetr}
\bibliography{Bakery}{}

\end{document}